# Operational Experience with the CMS Pixel Detector

**János Karancsi**[a,b] on behalf of the CMS Collaboration

[a] *Institute for Nuclear Research, Hungarian Academy of Sciences*
*H-4026 Debrecen, Bem tér 18/c, Hungary*

[b] *Institute of Experimental Physics, University of Debrecen*
*H-4026 Debrecen, Bem tér 18/a, Hungary*
*E-mail*: `janos.karancsi@cern.ch`

ABSTRACT: In the first LHC running period the CMS-pixel detector had to face various operational challenges and had to adapt to the rapidly changing beam conditions. In order to maximize the physics potential and the quality of the data, online and offline calibrations were performed on a regular basis. The detector performed excellently with an average hit efficiency above 99% for all layers and disks. In this contribution the operational challenges of the silicon pixel detector in the first LHC run and the current long shutdown are summarized and the expectations for 2015 are discussed.

# Contents

## 1. Introduction

The Compact Muon Solenoid (CMS) detector is a multi-purpose apparatus that operates at the Large Hadron Collider (LHC) at CERN [1]. At the center of its tracking system lies a hybrid pixel detector. The primary purpose of the detector is to provide high precision measurements of charged particle trajectories very close to the interaction point.

The sensors consist of high dose n-implants inserted into a high resistance n-substrate with rectifying pn-junctions on their backside, and are connected with readout chips (ROCs) by indium bump-bonds. The total number of pixels is approximately 66 million, each with a cell size of $100 \times 150 \mu m^2$. There are all together 15840 ROCs located on three barrel layers and two endcap disks (Fig. 1).

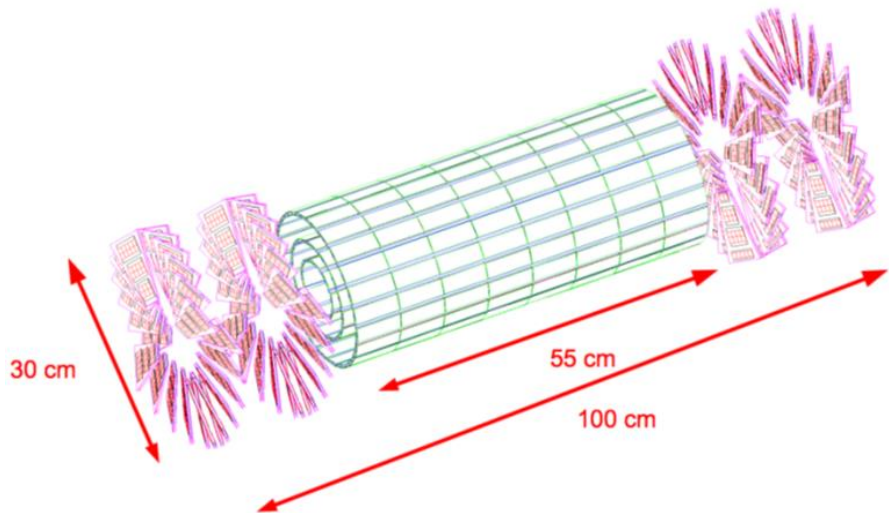

*Figure 1.* *Schematic view of the CMS Silicon Pixel detector. The interaction point is surrounded by three cylindrical barrel layers at radii of 4.4, 7.3 and 10.2 cm and two endcap disks on each side with modules between 6 and 15 cm from the beam axis.*

Many electronics components, such as power boards and electronics for optical readout, are located on the service structures. The Analogue Opto-Hybrid (AOH) converts the analogue electrical signal to an optical one. Optical signals are sent to the Front End Driver (FED) located outside the detector cavern, and then digitized by an Analog to Digital Converter (ADC). The calibration of these devices is discussed in Section 3.

## 2. Operational experiences

### 2.1 Challenges during Run I

The Large Hadron Collider has been in operation between 2008 and the beginning of 2013 which is known as Run I. By the end of 2012 the instantaneous luminosity of proton-proton collisions reached $7.7 \times 10^{32}$ cm$^{-2}$s$^{-1}$ and the average number of inelastic interactions (pile-up) was around 35. One of the main challenges was the precise reconstruction of primary and secondary vertices for which the detector required to maintain high efficiency and spatial resolution. The detector operated very well under these circumstances which was achieved with the continuous efforts of the operations team with frequent calibrations. This will be discussed further in Section 3.

Another important consequence of the high particle flux is the increased number of single event upsets (SEUs) that happen due to the ionization caused by a charged particle traversing the detector, upon which a bit is flipped in a logical element of the read-out electronics, causing an unwanted behavior. This may cause a temporary misbehavior of single pixels, readout chips or even whole modules. There were various recovery algorithms implemented in order to recover the silent components by reprogramming them in case a module was affected [2].

Ever since its installation the detector had to face various operational challenges that caused a permanent loss of modules. Most of the problems for the barrel pixels (BPix) appeared right after the installation, mostly due to broken wire bonds. Also, by the end of Run I seven modules with an old ROC design had to be reset frequently due to a rare readout error that increased with higher occupancy. This in turn caused an excessive downtime and consequently these modules had to be disabled.

A large fraction of the forward pixels (FPix) was removed, repaired and reinstalled in 2008. Right at the start of the LHC slow readout channels appeared on the FPix, upon which the analog signals could not be decoded correctly. Some of these channels was recovered successfully, but 3.6% of the FPix modules had to be disabled during data taking.

### 2.2 Challenges during Long Shutdown I

One of the main tasks of the Pixel team during the first long shutdown (LS1) of the LHC, between 2013 and 2014, was to repair or replace the bad components of the detector. By the end of Run I 19 modules (2.3%) of the BPix were not operational. Eight modules located on the outer face of the outmost layer and two AOHs have been successfully substituted. There is an ongoing effort to try to repair or replace all the remaining bad modules by the end of the shutdown.

A total of 7.8% of the FPix was not operational by the end of Run I. Most of the problems arose due to effectively unplugged or poorly connected AOH flex cables. The latter most likely caused the appearance of the previously mentioned slow channels with the distorted analogue signal. The flex cables have been replaced, reconnected and secured properly. The rest of the problematic modules have been replaced except for a single readout chip, which is located on a not easily accessible place, therefore 99.9% of the FPix will be fully operational for the next running period.

The pixel detector was operated with a $C_6F_{14}$ coolant temperature of 7.4°C between 2008 and 2011, and was cooled to 0°C in 2012. In order to avoid reverse annealing, limit the impact of radiation damage and to minimize leakage current the detector needs to be cooled well below 0°C [3], therefore the detector will operate at -15°C from 2015. In order to achieve this temperature safely a Tracker wide sealing was implemented in 2014 in order to ensure minimal humidity levels. The flow of dry gas into the tracker volume was increased and a new safety system is in place that shuts down the detector safely in case a sudden increase of temperature, electric current or humidity is detected. The full recalibration of the detector is also required to accommodate the new temperature.

### 3. Calibrations

In the commissioning period and throughout running the different components of the detector need to be calibrated in order to ensure fully efficient and reliable data collection. The analogue signal (Fig. 2) consists of data packets that contain the information about each pixel hit. The pixel coordinates are encoded by using a six-level address scheme, while the charge information is contained in the analogue pulse height of the last clock cycle [4].

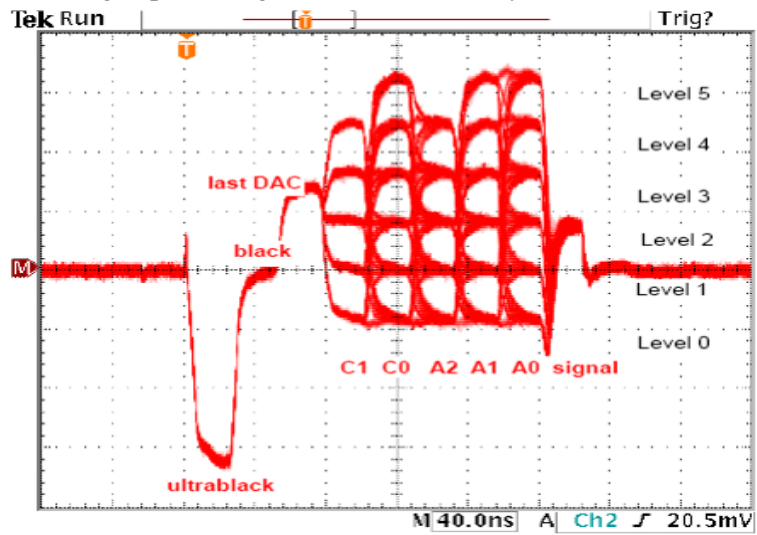

*Figure 2: Oscilloscope view of the analogue signal for a pixel hit.*

The offset of the optical receiver in the Front End Driver (FED) needs to be set such that the baseline level of the converted analogue signal is at the middle of the ADC range. This calibration is sensitive to temperature changes and needs to be done frequently, but small adjustments are done automatically to correct for small fluctuations. The address level calibration, which is usually done after the baselines are set, ensures that the ADC values corresponding to each address level are well separated and pixel coordinates can be decoded correctly.

Another set of calibrations needs to be done every few months due to the aging of the detector. One of them is the threshold calibration for which electric charges are injected through a circuit on the readout chip and the efficiency is measured. A scan is done for each pixel in units of $V_{cal}$ which corresponds to 65.5 electrons. The efficiency is measured and the threshold is derived from a fitted turn-on curve (Fig 3). The measurement [5] is then repeated in an iterative way for various ROC threshold settings in order to minimise the thresholds.

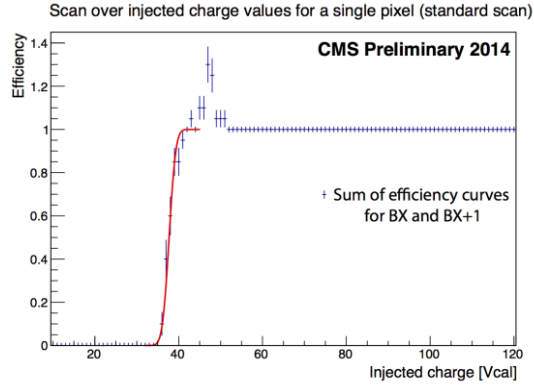

*Figure 3: Sum of in- and out-of-time charge collection efficiencies as a function of injected charge.*

During a gain calibration the ADC response of the optical converter to the incoming analogue signal is measured as a function of injected charges ($V_{cal}$). The gain and the pedestal, which is used during the offline reconstruction of clusters, is then extracted from a linear fit in the region well below the saturation point.

Usually once per year, at the beginning of data taking the timing calibration of the detector is done in which the pixel detector clock is synchronized with the LHC collisions. A scan is done for various settings of the relative phase delay with respect to the LHC clock. A delay is chosen which ensures the collection of both the high pixel charges and giving enough time for most of the smaller charge pulses to cross the readout threshold in the same 25ns readout window. The goal is to maximise both the hit- and the charge collection efficiency (Fig. 4) and to allow the usage of smaller thresholds which then improves the hit resolution. Therefore a point is chosen near the end of the efficiency plateau.

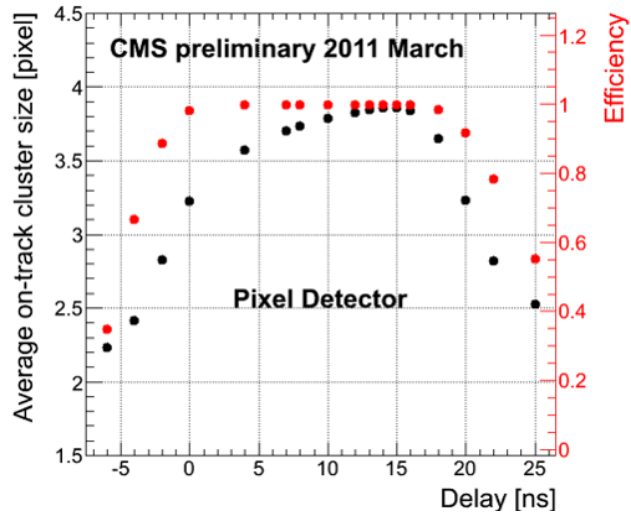

*Figure 4: Average cluster size and hit efficiency vs. time delay with respect to the LHC clock.*

## 4. Detector performance

The hit efficiency is an important parameter to monitor throughout data taking. It is defined as the probability to find a cluster along a charged particle trajectory. Nonfunctioning modules or those that were affected by a single event upset are excluded from the measurement. The method is explained in more detail in [6]. In 2012 the average efficiency of the detector was above 99% for all layers and disks (Fig. 5). Compared to the outer part of the detector, the innermost layer, which is affected by a larger particle flux, experienced a larger efficiency loss. This is an expected behavior, due to the limited buffer size in the current readout chip [7].

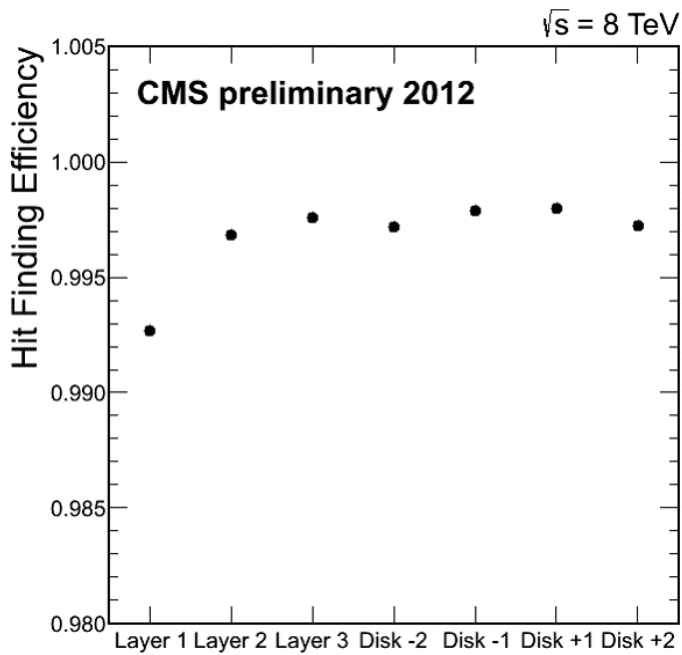

*Figure 5:* *The average hit efficiency for all barrel layers and endcap disks in 2012.*

## 5. Plans for 2015

In 2015, the LHC will provide collisions initially with a 50ns bunch spacing with similar instantaneous luminosities as in 2012 but with a center-of-mass energy of 13 TeV. Then the machine will switch to 25ns operation and further increase the rate of collisions to a planned peak instantaneous luminosity of $\sim 1.5 \times 10^{34}$ cm$^{-2}$s$^{-1}$ and an expected average number of inelastic proton-proton interactions of around 45.

In Run I the center of the pixel detector was slightly shifted from the beam axis by a few millimeters. During the current shutdown a new, thinner beam pipe has been installed that allows the current detector to be better centered and to age more uniformly. It also gives enough space to fit the Phase-I Upgrade detector that is going to be installed during the 2016/2017 technical stop. It has four barrel layers and three endcap disks and utilizes an improved version of the current readout chip with digital output.

A so-called pilot system has also been installed. It includes four Phase-1 modules on two forward half disks, DC-DC converters and new electronics for communication and optical readout in the service structure. The system will be thoroughly tested for functionality in the upcoming LHC run.

## 6. Conclusion

In the first running period, the LHC provided collisions with an unprecedented luminosity which posed various operational challenges under which the CMS pixel detector performed excellently with an average hit efficiency above 99% for all barrel layers and forward disks. This performance could only be achieved with a continuous effort in calibrations and online operations to adapt to the rapidly changing beam conditions.

By the end of the first LHC run 2.3% of the BPix and 7.8% of the FPix was not operational. Meanwhile 99.9% of the FPix is again operational and there is an undergoing effort to repair the barrel similarly.

In the next data taking period both the pile-up and the instantaneous luminosity is expected to increase which will require more frequent calibrations and new reconstruction techniques. The detector was prepared to run at a temperature of -15 °C by providing a new sealing and other protective measures in order to minimize humidity in the tracker volume and to protect the detector from aging rapidly. A new pilot system utilizing the Phase-I upgrade modules has been installed and will be thoroughly tested in the second LHC run.

## Acknowledgments

The author wishes to thank Viktor Veszpremi and Danek Kotlinski for their help in the preparation of this paper. This work is supported by the Hungarian Scientific Research Fund (OTKA K 109803) and the SCOPES Joint Research Project of the Swiss National Science Foundation.